\documentclass{iopart} \usepackage{graphicx}
\expandafter\let\csname equation*\endcsname\relax
\expandafter\let\csname endequation*\endcsname\relax
\usepackage{amsmath} \usepackage{amsfonts} \usepackage{multirow}
\newcommand{\ChEFT}{$\chi$EFT}
\begin{document}
\title{Statistical uncertainties of a chiral interaction at next-to-next-to leading order}
\author{A. Ekstr\"{o}m$^1$\footnote{Present address: Department of Fundamental
Physics, Chalmers University of Technology, SE-412 96 G\"{o}teborg, Sweden},  B.
D. Carlsson$^{2,1}$, K. A. Wendt$^{3,4}$, \\ C. Forss\'en$^2$, M.
Hjorth-Jensen$^{1,5}$,
R. Machleidt$^{6}$, \\and S. M. Wild$^{7}$}
\address{$^1$Department of Physics and Center of Mathematics for Applications, University of Oslo, N-0316 Oslo, Norway}
\address{$^2$Department of Fundamental Physics, Chalmers University of Technology, SE-412 96 G\"{o}teborg, Sweden }
\address{$^3$Department of Physics and Astronomy, University of Tennessee, Knoxville, TN 37996, USA}
\address{$^4$Physics Division, Oak Ridge National Laboratory, Oak Ridge, Tennessee 37831, USA}
\address{$^5$National Superconducting Cyclotron Laboratory and Department of Physics and Astronomy, Michigan State University, East Lansing, MI 48824, USA}
\address{$^6$Department of Physics, University of Idaho, Moscow, ID 83844, USA}
\address{$^7$Mathematics and Computer Science Division, Argonne National Laboratory, Argonne, IL 60439, USA}
\ead{jaeks@fys.uio.no}
\begin{abstract}
We have quantified the statistical uncertainties of the low-energy
coupling-constants (LECs) of an optimized nucleon-nucleon (NN)
interaction from chiral effective field theory (\ChEFT) at
next-to-next-to-leading order (NNLO). In addition, we have propagated
the impact of the uncertainties of the LECs to two-nucleon scattering
phase shifts, effective range parameters, and deuteron observables.
\end{abstract}
\submitto{\JPG}
\maketitle
\section{Introduction}
Chiral effective field theory (\ChEFT~\cite{EHM09,ME11}) defines the
current paradigm in the theoretical description of the interaction
between the nucleons. The algebraic structure and pattern are governed
by an effective Lagrangian with an associated power counting scheme.
However, as in the case of any nuclear interaction, the numerical
values of a set of low-energy coupling-constants (LECs) will govern
the quantitative behavior of the interaction to a large extent. Thus,
for a precise and successful modeling of the atomic nucleus, the
numerical values of the LECs must be well-constrained. In the
nucleon-nucleon (NN) sector, this is usually achieved by
confronting the LECs with data from scattering experiments. The pool
of fit observables consists of several thousands of measured
proton-proton ($pp$) and neutron-proton ($np$) scattering
cross-sections below the pion-production threshold at 290 MeV, and
sometimes also the measured deuteron properties. This constitutes an
extensive nonlinear optimization problem that is best tackled using
mathematical optimization algorithms~\cite{Ekstrom2013}. The
experimental scattering data that constrain the microscopic
interactions have well-defined statistical uncertainties. These
uncertainties must propagate to the LECs of the nuclear interaction, and
therefore also to any subsequently computed many-body observables.

In many cases the numerical solution of the many-body
Schr\"odinger equation is no longer the bottleneck in nuclear modeling;
see, e.g., Ref.~\cite{Binder2013}. Instead, it is the quality of the underlying
nuclear Hamiltonian that is responsible for observed discrepancies
between data and theory. In relation to this, it is important
to note that theoretical results from nuclear models are seldomly
reported with proper uncertainty estimates. However, the theoretical uncertainties
and error estimates of models are increasingly acknowledged as vital for 
making further theoretical developments; see, e.g., Ref.~\cite{Witeks_guide}. 

In this paper we address the statistical uncertainties
of the LECs of a two-nucleon interaction from $\chi$EFT. This
provides a first step toward a reliable extraction of the statistical
and systematic uncertainties of nuclear many-body observables
computed with \textit{ab initio} models. We present a mathematically
optimized two-nucleon interaction at next-to-next-to-leading order
(NNLO). From a statistical perspective, we can view the optimization
problem as a nonlinear regression problem, and therefore easily
extend the analysis to investigate the statistical constraints on the
LECs of this optimized interaction.

\section{Nuclear forces from chiral effective field theory}
From a historical perspective, it is interesting to see how nuclear
physicists have returned to Yukawa's idea~\cite{Yukawa}, at least
conceptually, but this time included the important notion of a
spontaneously broken chiral symmetry in the u-d quark sector of
quantum chromodynamics (QCD). As a consequence, pions are emerging as
pseudo-Goldstone bosons, whose interactions vanish in the low-momentum
limit. A chiral nuclear force is therefore grounded in the symmetries
of QCD and uses only nucleons and pions as degrees of freedom. It is
well known that QCD is nonperturbative at the energy scales relevant
for nuclear structure. A perturbative expansion of the nuclear force
is instead constructed from an effective field theory, using the
notion of a separation of scales~\cite{EHM09,ME11}. This approach
boils down to taking the most general nuclear-force Lagrangian,
obeying the symmetries of QCD, and subsequently expanding it in terms
of ($Q/\Lambda_{x}$) where $Q$ is a soft momentum-scale, typically the
pion mass or momentum, and $\Lambda_x$ sets the hard momentum-scale,
typically on the order of the rho-meson mass. It is equipped with a
power-counting scheme that governs the structure of the order-by-order
expansion. At each chiral order $\nu$, a certain number of new LECs
appear. These are related to either the long-ranged and Yukawa-like
pion-nucleon ($\pi-N$) interaction or to the short-ranged, and
unresolved, contact interaction. In relation to this, a long-standing
problem in nuclear physics has been to systematically construct two-
and higher-order nucleon interactions within a common framework. This
is remedied in $\chi$EFT, where two- and higher-order interactions are
borne out of the same perturbative expansion.

It was shown in
Ref.~\cite{Machleidt2003} that it is necessary to include terms in the
chiral expansion up to next-to-next-to-next-to-leading order (N3LO)
(i.e., $\nu=4$), in order for the interaction to quantitatively reproduce
the combined neutron-proton ($np$) and proton-proton ($pp$) scattering
data with laboratory scattering energies up to
$T_{\textnormal{LAB}}=290$ MeV. This so-called Idaho-N3LO potential is
widely used in various calculations of low-energy nuclear properties
such as binding energies, radii, and excited spectra. The number
of LECs grows with increasing chiral order, and at N3LO there are
34 LECs in total. Since these LECs need to be determined from experimental
data, this poses a rather difficult optimization problem. It was shown
in Ref.~\cite{Ekstrom2013}, that mathematical optimization algorithms
can be applied successfully in order to aid in the construction of
chiral potentials.

\begin{figure}
\center
\includegraphics[width=0.70\textwidth]{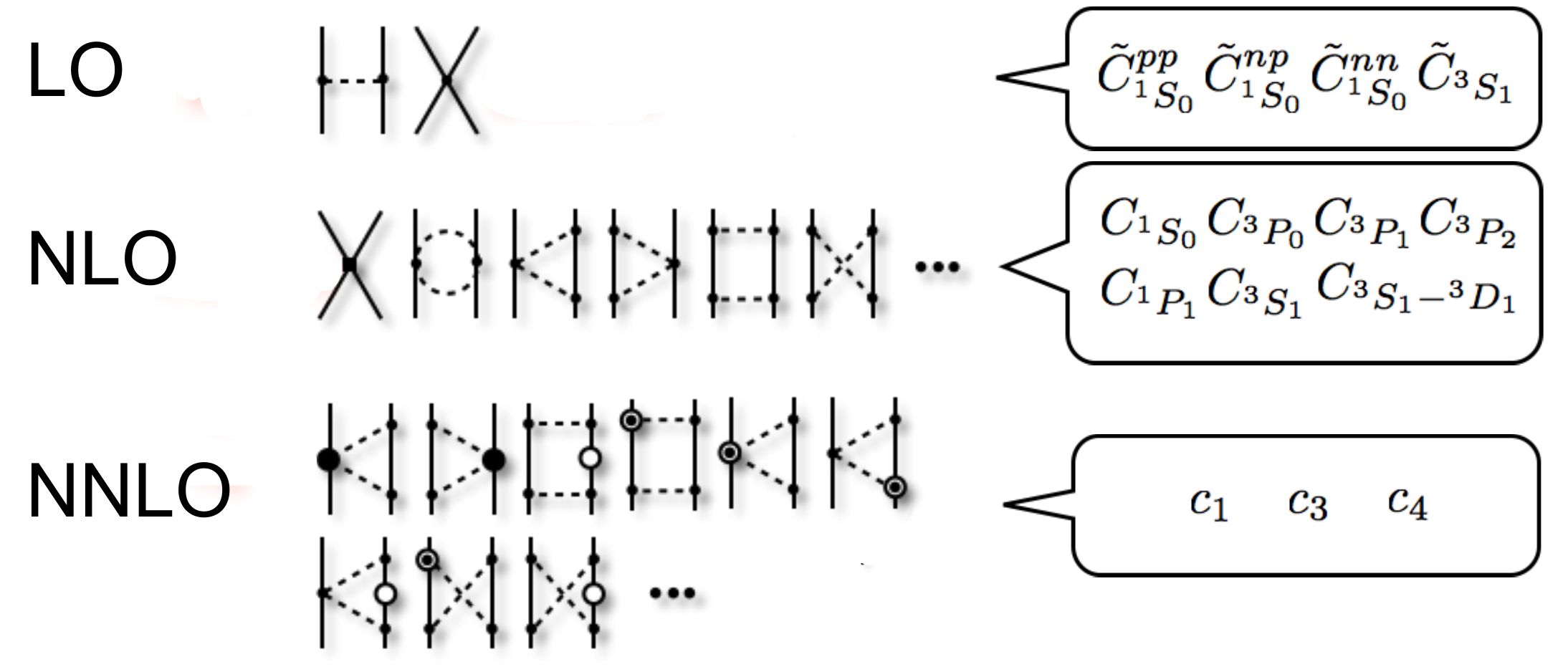}
\caption{The hierarchy of the nuclear force in \ChEFT, and in particular
  the two-nucleon sector up to NNLO. Solid lines indicate nucleons and
  dashed lines pions. The boxes on the right indicate the number and
  notation of the new LECs introduced at each chiral order. The
  $\tilde{C}$ and $C$ parameters belong to the two-nucleon
  contact-potential at LO and NLO, respectively. The $c_{i}'s$
  describe the long-ranged pion-mediated physics. Note that the
  isospin-breaking property of the leading-order $^{1}S_0$
  coupling-constant is a sub-leading effect. In relation to this,
  there exists no neutron-neutron scattering data. Thus the
  uncertainty of the $\tilde{C}_{^1S_0}^{nn}$ contact term will remain
  undetermined in this work. See the text for further details. }
\label{fig:chiral_expansion}
\end{figure}

For the present investigation, we will 
remain at the NNLO level and treat only the two-body interaction. At this order,
the dimensionality of the optimization problem is tractable while the
potential is still capable of a reasonable description of low-energy
nuclear physics~\cite{Ekstrom2013}. The details of the NNLO chiral
two-nucleon force that we study here are given in Ref.~\cite{ME11}. In
brief, we use the Weinberg power counting and regulate the potential
with a cutoff $\Lambda=500$ MeV in the solution of the
Lippmann-Schwinger equation. The two-pion exchange-potential is
renormalized using the spectral function renormalization prescription
with a cutoff $\Lambda_{\textnormal{SFR}}=700$
MeV~\cite{Epelbaum2004}. The familiar Yukawa-type one-pion
exchange-force enters at leading-order (LO) together with a simple
contact potential that will parameterize the $^1S_0$ and $^3S_1$
partial-waves of the interaction. The sub-leading two-nucleon
interaction NLO introduces a more sophisticated contact potential that
breaks the isospin of the LO contacts and parameterizes also the
deuteron channel and the $P-$waves. At NLO the leading two-pion
exchange also appears. The NNLO contribution only impacts the $\pi-N$
sector in terms of certain relativistic corrections and additional
two-pion exchanges that are proportional to the LECs $c_1,c_3$, and
$c_4$. In total, there are 14 LECs in the two-nucleon sector up to and
including NNLO. In summary, the various orders of the irreducible
graphs that define the NNLO potential are given by
(cf. \Fref{fig:chiral_expansion})
\begin{eqnarray}
V_{\rm LO} & = & 
V_{\rm ct}^{(0)} + 
V_{1\pi}^{(0)} 
\label{eq_VLO}
\\
V_{\rm NLO} & = & V_{\rm LO} +
V_{\rm ct}^{(2)} + 
V_{1\pi}^{(2)} +
V_{2\pi}^{(2)} 
\label{eq_VNLO}
\\
V_{\rm NNLO} & = & V_{\rm NLO} +
V_{1\pi}^{(3)} + 
V_{2\pi}^{(3)},  
\label{eq_VNNLO}
\end{eqnarray}
where the superscript denotes the order $\nu$ of the low-momentum
expansion. Contact potentials carry the subscript ``ct'' and
pion-exchange potentials can be identified by an obvious subscript.

\section{Definition of the objective function}
The optimization problem consists of determining an optimal set of
values for the 13 LECs $\mathbf{x}=(\tilde{C}_{^1S_0}^{np},
\tilde{C}_{^1S_0}^{pp}, \ldots, c_1,c_3,c_4)$ of the NNLO interaction
such that the $np$- and the $pp$-scattering data are best
reproduced. Note that we will not optimize the charge-dependent
neutron-neutron LO contact $\tilde{C}_{^1S_0}^{nn}$ with the other
LECs, since there exists no $nn$ scattering data. Instead, this
contact is determined from the model-dependent extractions of the $nn$
effective range parameters. We do not include any other data types in
the optimization protocol either, such as the deuteron binding energy,
since we have a subsequent statistical analysis of the optimum in
mind. A mixture of data types would introduce an additional parameter
to determine the relative importance of, for example, cross sections
versus a binding energy. Note also that we keep the axial-vector coupling
constant $g_A=1.29$ to account for the Goldberger-Treiman discrepancy,
the weak-pion decay $f_{\pi}=92.4$ MeV, and all pion- and
nucleon-masses fixed. It was shown in Ref.~\cite{Ekstrom2013} that a
chiral potential at NNLO with $\Lambda=500$ MeV and
$\Lambda_{SFR}=700$ MeV can describe the two-nucleon scattering data
up to a laboratory scattering-energy $T_{LAB} \leq 125$ MeV. This is
therefore our cutoff in the experimental datapool used in the fit. We
also note that the optimized NNLO potential presented here differs
from NNLO$_{\rm opt}$~\cite{Ekstrom2013} in terms of the definition of
the objective function. The latter potential was optimized with
respect to a selected class of phase shifts and the deuteron binding
energy. In this study we wish to estimate the uncertainties of the
LECs, and for that reason we have constructed an objective function
that consists of the experimentally measured cross sections and their
associated experimental uncertainties.

The
experimental data is composed of $N_g$ groups of measurements, where
each group consists of $N_d$ measured cross sections
$\mathcal{O}_{g,d}^{e} \pm \sigma_{g,d}$ with a common normalization
constant $\nu_g$ and associated error $\sigma_{g,0}$. A group of data
originates from the same experiment. The normalization constant,
together with its uncertainty, represents the systematic uncertainty
of the measurement. For an absolute measurement, the normalization is
given by $\nu_g=1 \pm 0$. Usually, this means that the statistical and
systematic errors have been combined with $\sigma_{g,d}$, but certain
experiments are not normalized at all. Instead, only the angular- or
$T_{LAB}$-dependence of the cross section was determined. For these
groups of data, so-called ``floated data'', $\nu_g$ is solved for in the
optimization by minimizing the discrepancy between the model
prediction $\mathcal{O}_{g,d}^{t}$ and the experimental data points
$\mathcal{O}_{g,d}^{e}$. For practical purposes the normalization
error can be considered infinite in these cases, and will therefore
not contribute to $\chi^2$. Statistically, we seek to find the
minimum of $\chi^2$ with respect to variations of the (LEC) parameter
vector $\mathbf{x}$. Thus, we are after a minimizer of 
an objective function of the form
\begin{equation}
\chi^2(\mathbf{x}_{\mu}) = \underset{\mathbf{x}}{\min}\left\{\sum_{g=1}^{N_g}
\underset{\nu_g}{\min}\left[\sum_{d=1}^{N_d} \left(\frac{
\nu_g\mathcal{O}_{g,d}^{t}(\mathbf{x}) -
\mathcal{O}_{g,d}^{e}}{\sigma_{g,d}}\right)^2\right]+
\left(\frac{1-\nu_g}{\sigma_{g,0}}\right)^2\right\}.
\label{eq:objective_function}
\end{equation} 
The present dataset consists of $N_{obs}=1848$ measured data points
and $N_{\nu}=108$ normalization constants, out of which $N_{float}=11$
data sets are floated. Thus, there are
$N_{datum}=N_{obs}+N_{\nu}-N_{float} = 1945$ terms in the objective
function, and $N_{\nu}-N_{float}=97$ of them come from the
normalization of certain data groups. The total number of parameters
used in the optimization is $N_{par}=N_{float}+N_{NNLO} = 24$, where
$N_{NNLO}=13$ is the total number of varied parameters in the NNLO
potential. The number of degrees of freedom in the optimization
amounts to $N_{df}=N_{datum}-N_{par} = 1921$. In the
nuclear-interaction community, the quality of the potential is gauged
by the $\chi^2/N_{datum}$. However, for the statistical distributions
that are applied in the statistical analysis of the potential we will
use $N_{df}$. The present data set is based on the SM99 data
set~\cite{Machleidt2001}, with the exclusion of 7+25 datapoints, see
\Tref{tab:data}. Seven data points were removed based on the
$3\sigma$-rejection rule~\cite{Ber88} with respect to a CD-Bonn
prediction~\cite{Machleidt2001}. The origin of the discrepancy with
respect to the $3\sigma$-rejection is most likely due to slightly
different numerics in the computer codes that solves the two-nucleon
scattering problem. An additional 25 datapoints were removed due to
numerical issues when differentiating the Lippmann-Schwinger equation
of the $^3S_1-^3D_1$ coupled-channel in the bound-state cross-over
region around $17$ MeV. The removal of these latter 25 datapoints had
no significant impact on the solution of the optimization problem.
\begin{table}
\center
\caption{The experimental database consists of the SM99 database
  with the exclusion of the following data points.}
\label{tab:data}
\lineup
\begin{tabular}{@{}llll}
\br
$E_{\textnormal{LAB}}$ & $\theta_{\textnormal{CM}}$ & type & Ref. \\
\mr
7.6  & 90.6       & $np$ $P$ & \cite{Weisel92} \\
16.9 & all        & 15 $np$ $P$ & \cite{Tornow88} \\
16.9 & all        & 4 $np$ $P$ & \cite{Morris74}\\
17.0 & all        & 6 $np$ $P$ & \cite{Wilczynski84}\\
22.0 & 141.0      & $np$ $P$ & \cite{Wilczynski84} \\
29.9 & 164.9      & $np$ $\sigma$ & \cite{Fink90} \\
54.0 & --         & $np$ $\sigma_{\textnormal{tot}}$ & \cite{Lisowski82} \\
67.5 & 46.0       & $np$ $\sigma$ & \cite{Bersbach76} \\
97.7 & 46.0,71.4  & 2 $pp$ $P$ & \cite{Wigan68} \\
\mr
\end{tabular}
\end{table}

\section{The optimized NNLO potential}
We solve the optimization problem defined in
\Eref{eq:objective_function} using an optimization-routine called
POUNDerS \cite{SWCHAP14,tao-man}, which has also been applied to 
optimize nuclear energy
density functionals~\cite{Kortelainen2010}. The POUNDerS algorithm
does not rely on derivatives of the objective function with
respect to the model parameters. Instead, it solves a sequence of easier
subproblems based on locally fitting a collection of quadratic surfaces to the 
residual values in the objective function. The vector $\mathbf{x}_{\mu}$ of the 
LECs that
minimize the $\chi^2$ in (\ref{eq:objective_function}) is given in 
\Tref{tab:LECs}. The value of the
objective function at the minimum is $\chi^{2}(\mathbf{x}_{\mu}) =
2243.5$, which gives $\chi^{2}/N_{datum} = 1.15$ over the
$T_{\textnormal{LAB}}$-range 0-125 MeV. In detail, the
$\chi^2_{np}/N_{datum}$ in the bins $T_{\textnormal{LAB}}=0-35$ MeV
and $T_{\textnormal{LAB}}=35-125$ MeV are 0.87 and 1.24, respectively. 
Similarly, the
$\chi^2_{pp}/N_{datum}$ in these bins are 1.04 and 1.53, respectively. Thus, the
optimized NNLO potential quantitatively describes the $np+pp$
scattering data below $T_{\textnormal{LAB}} = 125$ MeV and is accurate
enough to model the bound-state properties of light- and medium-mass
nuclei.
\begin{table}
\center
\caption{Numerical values of the LECs for the optimized chiral NNLO
  potential. The optimum is defined by the $\mathbf{x}_{\mu}$
  vector. The standard-deviation $\sigma$ for each parameter, except
  $\tilde{C}_{^1S_0}^{nn}$, is given in the second column. A 95\%
  confidence interval is given in the third column. See text for
  details.}
\label{tab:LECs}
\lineup
\begin{tabular}{llllc}
\br
            LEC           & $x_{\mu}$ & 1$\sigma$ & $1\sigma/x_{\mu}$ [\%] & 95\% CI\\  
\mr
$c_1$                  &   \-0.919353                &       5.41$\cdot 10^{-2}$&          5.88 &   $[-1.008365, -0.830340]$  \\
$c_3$                  &    \-3.889839                &      1.68$\cdot 10^{-2}$&          0.43 &   $[-3.917492, -3.862185]$  \\
$c_4$                  &     4.307371                &       4.22$\cdot 10^{-2}$&          0.98 &   $[ 4.237956,  4.376786]$  \\
$\tilde{C}^{np}_{^1S_0}$ &  \-0.152151                   &   4.01$\cdot 10^{-4}$  &          0.26 & $[-0.152811, -0.151491]$     \\
$\tilde{C}^{pp}_{^1S_0}$ &  \-0.151363                   &   3.86$\cdot 10^{-4}$  &          0.25 & $[-0.151998, -0.150728]$     \\
$\tilde{C}^{nn}_{^1S_0}$ &  \-0.151804                   &   $-$  &          $-$ & $ -$ \\                                  
$\tilde{C}_{^3S_1}$     &   \-0.158482                 &     2.49$\cdot 10^{-4}$ &          0.16 &  $[-0.158891, -0.158073]$     \\
$C_{^1S_0}$             &    2.404311                 &      3.36$\cdot 10^{-3}$ &          0.14 &  $[ 2.398774,  2.409849]$     \\
$C_{^3P_0}$             &    1.235001                 &      9.51$\cdot 10^{-3}$ &          0.77 &  $[ 1.219344,  1.250658]$     \\
$C_{^1P_1}$             &    0.414829                 &      1.09$\cdot 10^{-2}$ &          2.62 &  $[ 0.396950,  0.432707]$     \\
$C_{^3P_1}$             &   \-0.770879                 &     7.17$\cdot 10^{-3}$ &          0.93 &  $[-0.782678, -0.759079]$     \\
$C_{^3S_1}$             &    0.927936                 &      3.12$\cdot 10^{-3}$ &          0.34 &  $[ 0.922809,  0.933064]$     \\
$C_{E_1}$              &     0.618754                 &      2.53$\cdot 10^{-3}$&          0.41  &  $[ 0.614584,  0.622925]$     \\
$C_{^3P_2}$             &   \-0.673469                 &     4.54$\cdot 10^{-3}$ &          0.67 &  $[-0.680941, -0.665996]$     \\
\br
\end{tabular}
\end{table}

\section{Statistical analysis}
\begin{figure}
\center
\includegraphics[width=0.9\textwidth]{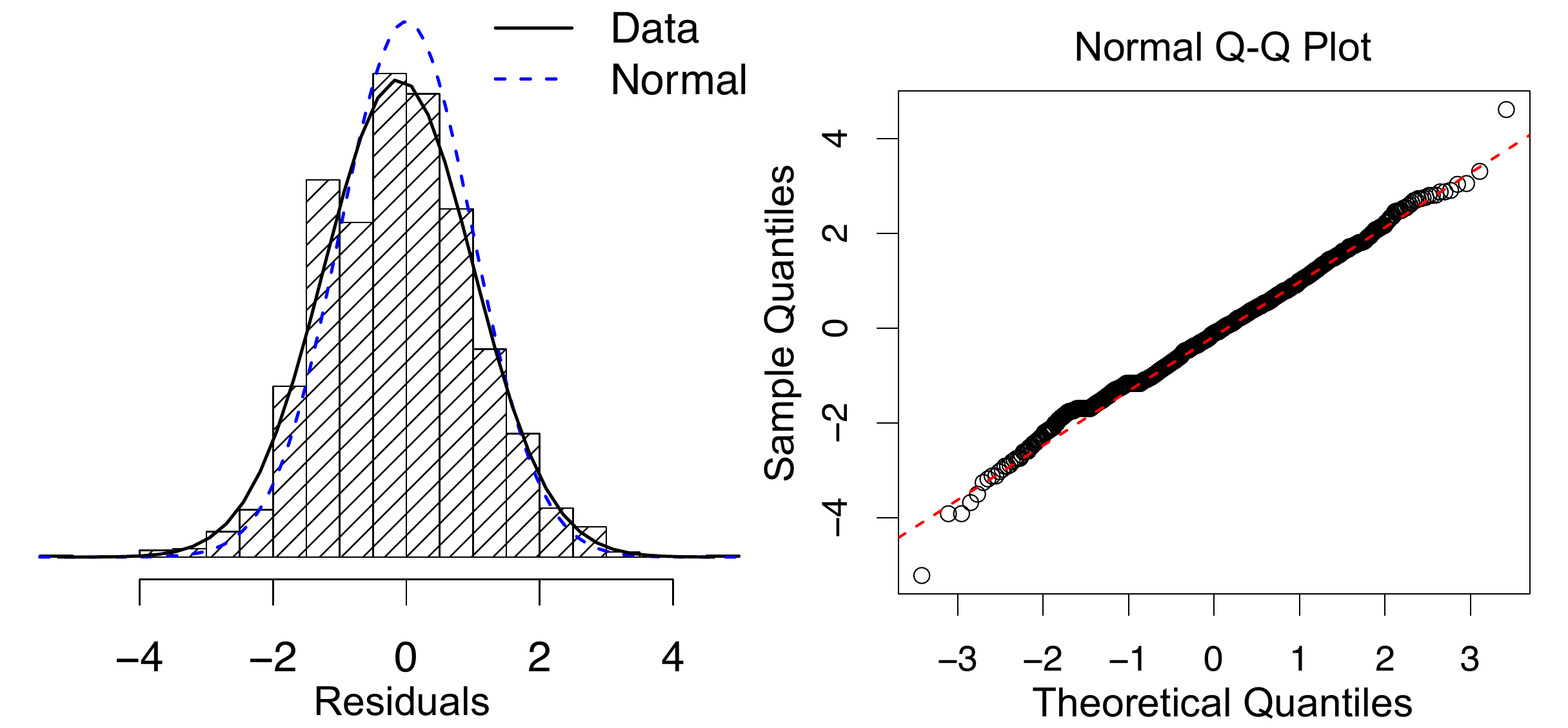}
\caption{(Left) The residuals of the objective function at
  $\mathbf{x}_{\mu}$ are fairly normally distributed. For comparison,
  a standard normal ($N(0,1)$) distribution function (blue dashed
  line) is also plotted. The skewness and excess kurtosis of the
  histogram of residuals are 0.06 and 0.37, respectively. A fitted
  normal distribution is indicated with a black solid line. (Right) A
  quantile-quantile plot that compares the normality of the residuals
  with the normal distribution.}
\label{fig:normal}
\end{figure}

We now turn our attention to the estimation of the statistical
uncertainties of the optimal LEC values $\mathbf{x}_{\mu}$. Formally, the
analysis in this section requires that the true errors in the objective
function are normally distributed and independent. Although this is a rather
strong requirement and hard to fulfill rigorously,
\Fref{fig:normal} illustrates that the residual errors for our 
$\mathbf{x}_{\mu}$ are approximately normal.
Consequently, we conduct the corresponding statistical
analysis in order to gain more insights into the NNLO potential. The
statistical uncertainty in $\mathbf{x}_{\mu}$ will propagate to an
uncertainty in the response of any many-body model based on this NNLO
potential. As a first step, we have carried out a preliminary study on
the extraction of statistical uncertainties on the LECs and propagated
the errors to a selected set of deuteron observables, and quantified
the uncertainties in the scattering phase-shifts as well as the
effective range parameters for the $^{1}S_0$ channel.

\subsection{Theoretical model uncertainties}
In a statistical setting, the optimization of the objective function
also represents a nonlinear regression problem. We can view the input
vector $\mathbf{x}_{\mu}$ as a normally distributed random vector
whose variance we seek. In order to determine the covariances of the
model parameters, we follow \cite{Witeks_guide} and perform a Taylor expansion 
of the objective function $\chi^2$ around the mean vector $\mathbf{x}_{\mu}$:
\begin{equation}
\chi^2(\mathbf{x}) - \chi^2(\mathbf{x}_{\mu}) \approx
\frac{1}{2}\sum_{i,j=1}^{N_p} \left.\frac{\partial^2\chi^2}{\partial
  x_i \partial
  x_j}\right|_{\mathbf{x}=\mathbf{x}_{\mu}}(x_i-x_{{\mu},i})(x_j-x_{{\mu},j}).
\end{equation}
The linear term drops out since $\mathbf{x}_{\mu}$ is a minimum. The
approximations that we make in this analysis depend on staying
sufficiently close to this point in parameter space. The scaled Hessian
$\mathbf{H}$, with elements
$H_{ij}=\frac{1}{2}\left.\frac{\partial^2\chi^2}{\partial x_i \partial
  x_j}\right|_{\mathbf{x}=\mathbf{x}_{\mu}}$, is related to the
covariance matrix of the model parameters as
\begin{equation}
\mathbf{\Sigma} = \frac{\chi^2}{N_{df}}\mathbf{H}^{-1}.
\end{equation}
In this analysis, we obtained the Hessian matrix $\mathbf{H}$ from a
bivariate spline in the vicinity of the optimum.
The square root of the
diagonal elements of the covariance matrix
$\mathbf{\Sigma}=\textnormal{Cov}(\mathbf{x}_i,\mathbf{x}_j)$ are
interpreted as the errors of the LECs: $\sigma_{i} =
\sqrt{\Sigma_{ii}}$. In addition to extracting the errors, we deduce a
95\% confidence interval (CI) for every parameter according to the
$t$-distribution
\begin{equation}
x_{\mu,i} - \sigma_i \cdot t_{N_{df},\alpha/2} \leq x_i \leq x_{\mu,i} + 
\sigma_i \cdot t_{N_{df},\alpha/2},
\end{equation}
where $\alpha=0.05$. The CI can be interpreted as the range of
acceptable values for constructing the current NNLO potential. In
\Tref{tab:LECs} we list statistical uncertainties of the LECs together
with 95\% CIs. In summary, the uncertainties are at the 1\%-level, but
two LECs, $c_1$ and $C_{^1P_1}$, stand out and have relatively large
errors. It is interesting that $c_1$ exhibits an approximately five
times larger uncertainty compared to the other $\pi-N$ LECs. This is
in qualitative agreement with the similar analysis of
Ref.~\cite{Navarro14}, albeit using a different representation of the
short-ranged interaction. The different representation of the contact
potential, or simply the fact that we are at a completely different
optimum, could be the reasons for the drastically smaller
uncertainties that we observe here. In general, the $S-$wave contacts
are slightly more precise compared to the $P-$wave contacts. This is
perhaps not so surprising, since the $P-$wave phase-shifts are
difficult to reproduce at NNLO~\cite{Ekstrom2013}. The isoscalar
$^{1}P_1$-contact is determined only by the $np$-data, which is
overall much less precise compared to the $pp$-data.

\begin{figure*}
\center
\includegraphics[width=0.7\textwidth]{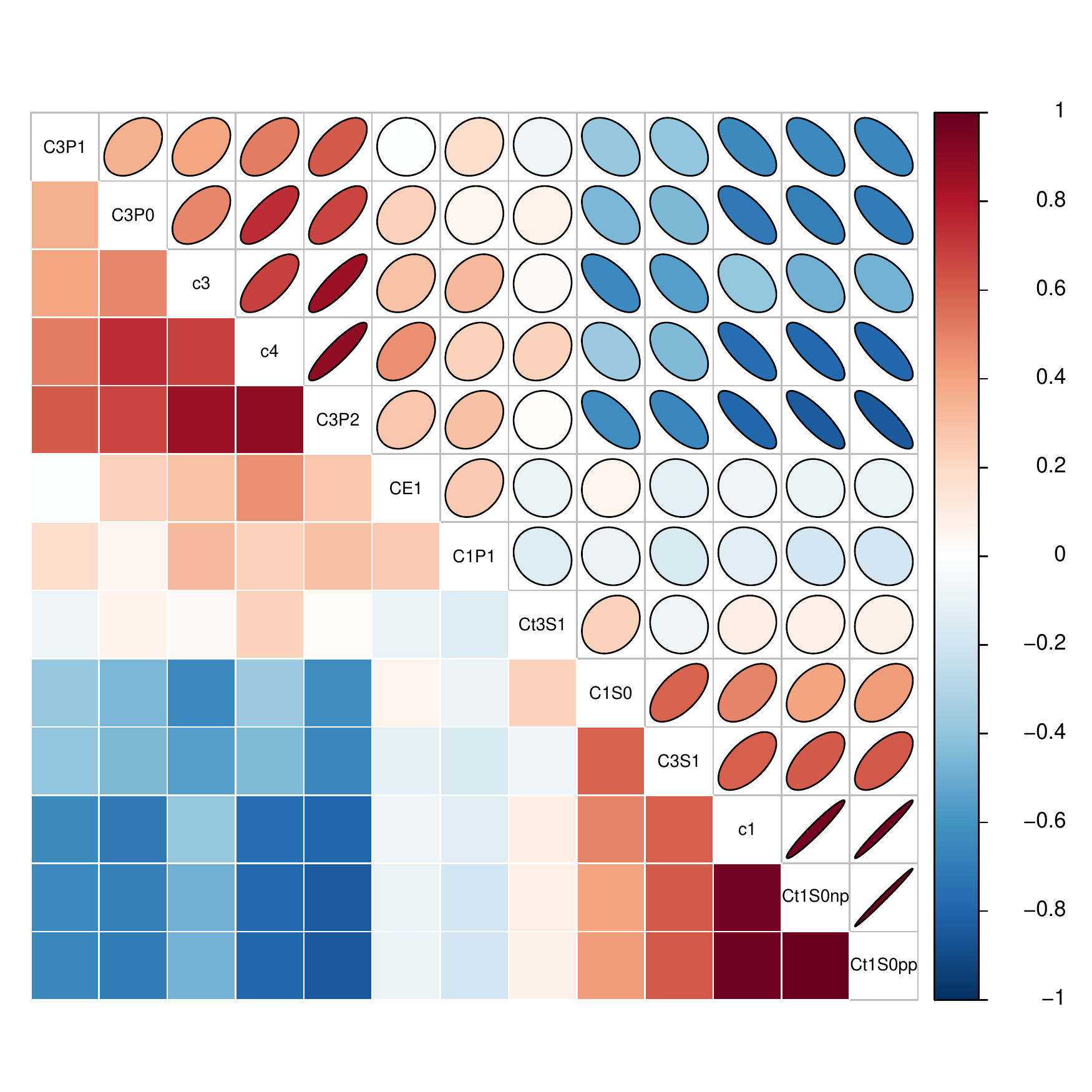}
\caption{Graphical representation of the correlation matrix for the
  current NNLO potential. Maximum correlation and anti-correlation is
  indicated with blue and red colors, respecively. The upper triangle
  contains all the correlation ellipses. The matrix has been grouped
  in blocks of similar correlation. The block structure of the $S-$
  and $P-$wave correlations is clearly visible. See the text for
  details.}
\label{fig:correlation_matrix}
\end{figure*}

\begin{figure*}
\center
\includegraphics[width=0.8\textwidth]{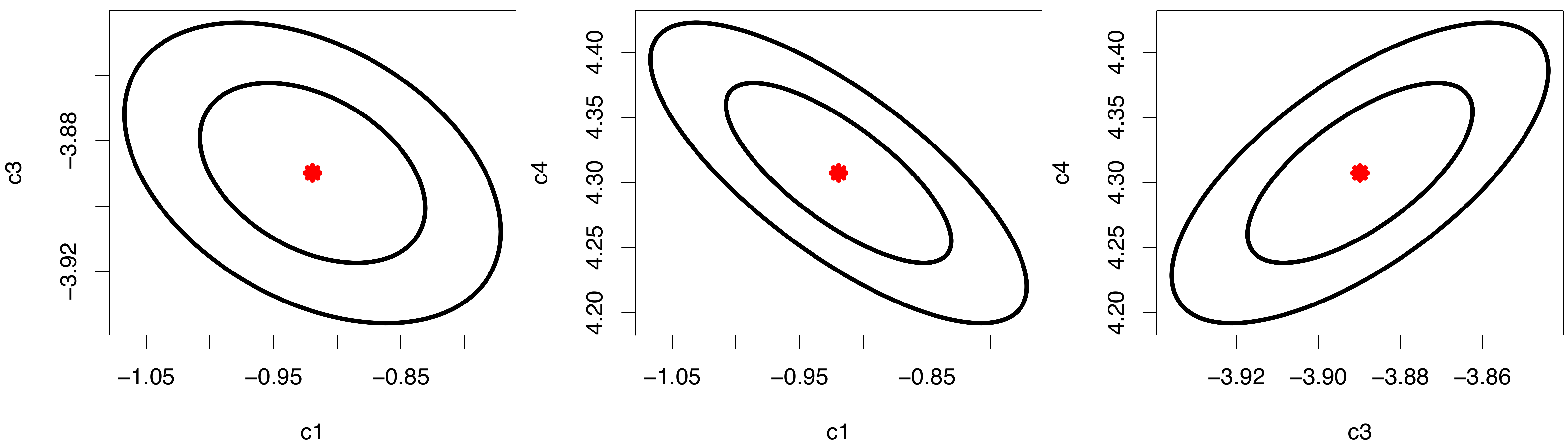}
\caption{Correlations between the $c_1,c_3,$ and $c_4$ LECs at the
  current NNLO optimum. The inner and outer contours trace the
  95\% and 99\% confidence ellipses, respectively.}
\label{fig:piN_correlations}
\end{figure*}

The correlations between the LECs provide further
insights into the behavior of the current model. Equipped with the
covariance matrix $\mathbf{\Sigma}$ of the LECs, it is a simple task
to extract the correlation coefficient between the optimal parameters $x_{\mu,i}$
and $x_{\mu,j}$ as
\begin{equation}
R_{ij}=\frac{\Sigma_{ij}}{\sqrt{\Sigma_{ii}\Sigma_{jj}}}.
\end{equation}
The correlation matrix $\mathbf{R}$ is given in \Tref{tab:correlation_matrix}
\begin{table}
\center
\caption{Correlation matrix for the NNLO potential $\mathbf{x}_{\mu}$.}
\footnotesize 
\label{tab:correlation_matrix}
\lineup
\begin{tabular}{@{}llllllllllllll}
\br
                       & $c_1$&$c_3$&$c_4$&$\tilde{C}^{np}_{^1S_0}$&$\tilde{C}^{pp}_{^1S_0}$&$\tilde{C}_{^3S_1}$&   $C_{^1S_0}$&$C_{^3P_0}$&$C_{^1P_1}$&$C_{^3P_1}$&$C_{^3S_1}$&$C_{E_1}$&$C_{^3P_2}$ \\  
\mr
$c_1$                  &  1.00   &          &            &           &          &            &          &           &           &           &            &          &       \\       
$c_3$                  & \-0.39   &   1.00   &            &           &          &            &          &           &           &           &            &          &       \\       
$c_4$                  & \-0.76   &   0.68   &    1.00    &           &          &            &          &           &           &           &            &          &       \\       
$\tilde{C}^{np}_{^1S_0}$ &  0.96   &  \-0.48   &   \-0.78    &   1.00    &          &            &          &           &           &           &            &          &       \\       
$\tilde{C}^{pp}_{^1S_0}$ &  0.98   &  \-0.48   &   \-0.79    &   0.99    &   1.00   &            &          &           &           &           &            &          &       \\       
$\tilde{C}_{^3S_1}$     &  0.09   &   0.03   &    0.23    &   0.08    &   0.07    &   1.00    &          &           &           &           &            &          &       \\       
$C_{^1S_0}$             &  0.49   &  \-0.64   &   \-0.37    &   0.40    &   0.42    &   0.23    &   1.00   &           &           &           &            &          &       \\       
$C_{^3P_0}$             & \-0.71   &   0.48   &    0.74    &  \-0.69    &  \-0.70    &   0.07    &  \-0.46   &    1.00   &           &           &            &          &       \\       
$C_{^1P_1}$             & \-0.12   &   0.32   &    0.22    &  \-0.19    &  \-0.19    &  \-0.14    &  \-0.08   &    0.06   &    1.00   &           &            &          &       \\       
$C_{^3P_1}$             & \-0.64   &   0.40   &    0.51    &  \-0.64    &  \-0.65    &  \-0.08    &  \-0.39   &    0.36   &    0.18   &    1.00   &            &          &       \\       
$C_{^3S_1}$             &  0.60   &  \-0.55   &   \-0.44    &   0.61    &   0.62    &  \-0.07    &   0.59   &   \-0.45   &   \-0.16   &   \-0.40   &    1.00   &           &       \\       
$C_{E_1}$              & \-0.07   &   0.29   &    0.46    &  \-0.09    &  \-0.10    &  \-0.09    &   0.06   &    0.23   &    0.26   &   \-0.01   &   \-0.11   &    1.00   &         \\       
$C_{^3P_2}$             & \-0.79   &   0.85   &    0.89    &  \-0.84    &  \-0.84    &   0.02    &  \-0.62   &    0.67   &    0.30   &    0.61   &   \-0.65   &    0.28   &    1.00 \\ 
\br
\end{tabular}
\end{table}
, and a graphical representation of $\mathbf{R}$ is shown in \Fref{fig:correlation_matrix}.
For this NNLO interaction, the strongest correlations occur between
$c_1$ and the charge-dependent $^{1}S_0$ contacts. There is also some
correlation between the three different $c_i$'s, see
\Fref{fig:piN_correlations}. Besides these, the correlation matrix
does not signal a strongly correlated set of parameters. In fact, $93
\%$ of the elements indicate a correlation $|R|<0.80$. Both
$c_1$/$c_3$ anti-correlate/correlate rather strongly with $c_4$,
at least compared to the correlation $R(c_1,c_3)$. It should be noted
that both $c_1$ and $c_3$ belong to the central part of the potential,
whereas $c_4$ belongs to the spin-spin and tensor parts of the
interaction. Thus, the observed correlations could be a weak
manifestation of the expected interplay between the central and the
tensor parts of the nuclear interaction~\cite{MachleidtANP}. Also, the
anti-correlation between $c_1$ and $c_3$ of the chiral two-pion
exchange potential was observed already in~\cite{Navarro14}. However,
the (weaker) correlations $R(c_1,c_4)$ and $R(c_3,c_4)$ have opposite
signs in the analysis of~\cite{Navarro14}. As mentioned, the LECs of
the contact potential of a chiral NNLO interaction enter only in $S-$
and $P-$waves. From the analysis it is clear that these two groups of
LECs mostly anti-correlate. Overall, this picture of a certain
partial-wave grouping is consistent with the observations made in a
similar analysis based on various coarse-grained $\delta$-shell
interactions~\cite{Navarro14,Navarro13_1,Navarro13_2}.

\section{Error propagation and uncertainties of the deuteron observables}
Equipped with statistical uncertainties of the LECs for the current
NNLO, we proceed with an error propagation of these to the deuteron
observables, $^{1}S_0$ effective range parameters, and selected
partial-wave phase-shifts.

In general, we are after the response $Y$
of a model $\mathcal{M}$ that depends on an input random vector
$\mathbf{X}$, and thus the response itself becomes a random
vector. For the present application, we can restrict the model to a
scalar-valued mathematical function $\mathcal{M}:\mathbb{R}^{M}
\rightarrow \mathbb{R}$. The response $Y$ is a random variable given
by
\begin{equation}
Y=\mathcal{M}(\mathbf{X}),
\end{equation}
and we seek the variance of the probability density function (pdf)
$f_{Y}(y)$, which will depend on the pdf of the input
$\mathbf{f}_{\mathbf{X}}(\mathbf{x})$. Here, we will use a
derivative-based approximation to the response variance
$\sigma_{Y}$. This will require the computation of as many derivatives
$\frac{\partial \mathcal{M}}{\partial x_i}$ as there are parameters
$x_i$, but will not produce a pdf $f_{Y}(y)$ for the response. Another
alternative for propagating the uncertainties of the LECs would be to
do a simple Monte Carlo sampling of a normally distributed nuclear
interaction. That is, to compute the observable of interest $N_{\rm
  MC}$ times using an interaction with a 
normally distributed set of LECs. We carried out such an analysis, and
the results are nearly identical to the computationally less expensive method
based on error propagation by means of perturbation.

\subsection{Error propagation by means of perturbation}
\begin{figure*}[h]
\center
\includegraphics[width=0.8\textwidth]{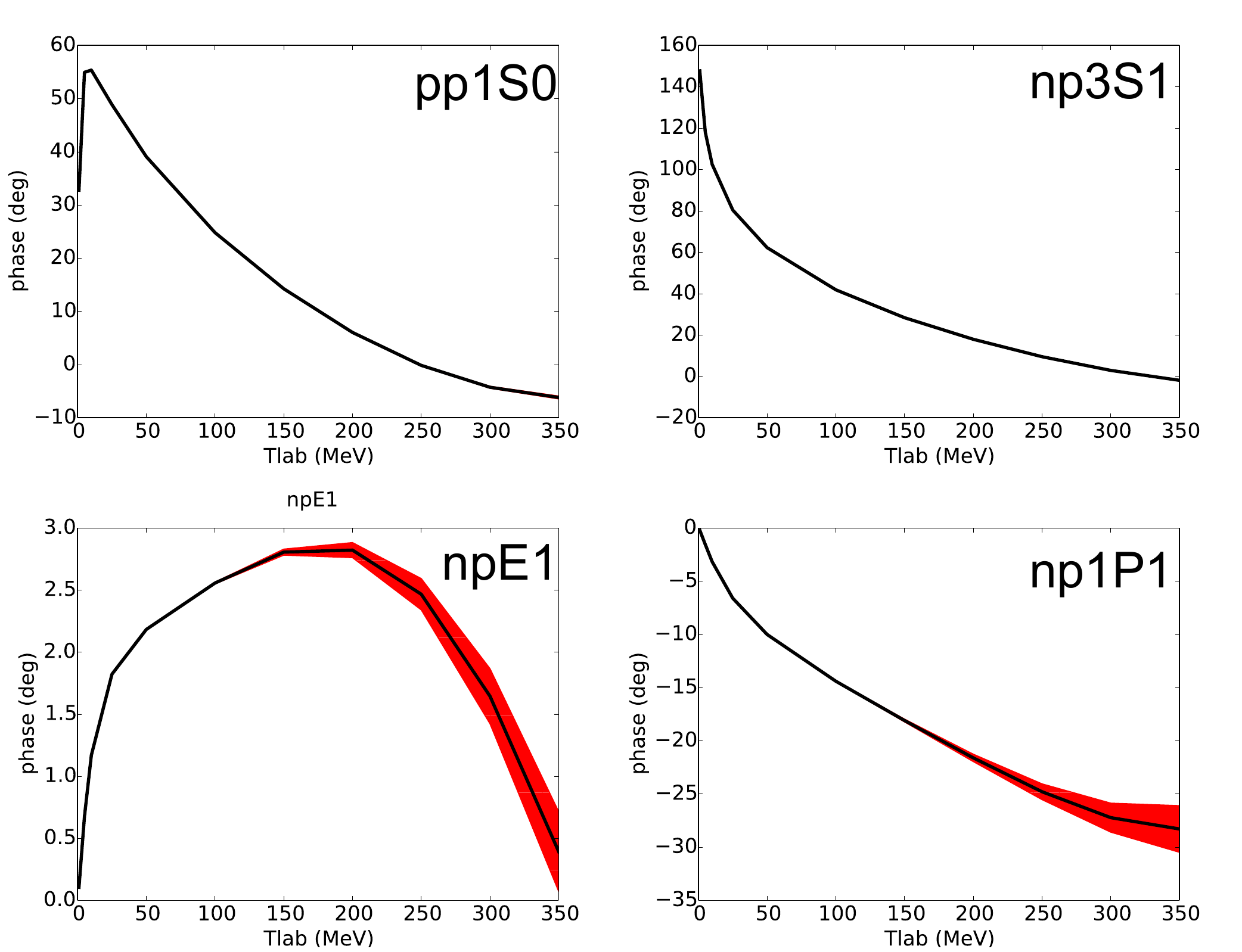}
\caption{Selected proton-proton and neutron-proton scattering phase
  shifts. The statistical $1\sigma$ uncertainties from the covariance
  matrix of the NNLO interaction are indicated with a red band. Only
  for the $npE1$ and $np1P1$ phase shifts are the statistical
  uncertainties visible when plotted on a scale set by the magnitude
  of the phase shift.}
\label{fig:phase_shifts}
\end{figure*}

A Taylor series expansion of the model $\mathcal{M}(\mathbf{x})$
around its mean value $\mathbf{x}_{\mu}$ 
is given by
\begin{align}
\begin{split}
Y = \mathcal{M}(\mathbf{x}) ={}& \mathcal{M}(\mathbf{x}_{\mu}) + \sum_{i=1}^N \left.\frac{\partial \mathcal{M}}{\partial x_i}\right|_{\mathbf{x}=\mathbf{x}_{\mu}} (x_i-x_{\mu,i}) \\
{}& + \frac{1}{2}\sum_{i,j=1}^{N} \left. \frac{\partial^2 \mathcal{M}}{\partial 
x_i \partial x_j} \right|_{\mathbf{x}=\mathbf{x}_{\mu}} 
(x_i-x_{\mu,i})(x_j-x_{\mu,j}) + \ldots .
\label{eq:model_expansion}
\end{split}
\end{align}
From the definition and linearity of the statistical expectation-value
operator $E[\cdot]$, and assuming that $\mathbf{x}_{\mu}$ is a
minimizer of $\mathcal{M}$ we have
\begin{align}
\begin{split}
E[Y] = E[\mathcal{M}(\mathbf{x})] \approx \mathcal{M}(\mathbf{x}_{\mu}) + 
\frac{1}{2}\sum_{i,j=1}^{N} \left. \frac{\partial^2 \mathcal{M}}{\partial x_i 
\partial x_j} \right|_{\mathbf{x}=\mathbf{x}_{\mu}} \textnormal{Cov}(x_i,x_j),
\end{split}
\end{align}
where we also identified the covariance $\textnormal{Cov}(x_i,x_j) = 
E[(x_i-x_{\mu,i})(x_j-x_{\mu,j})]$. The variance of the response is defined as
\begin{equation}
\textnormal{Var}[Y] = E[(Y-E[Y])^2].
\end{equation}
This can be approximated using the first-order term in the expansion
of $\mathcal{M}$ in \Eref{eq:model_expansion}
\begin{equation}
\textnormal{Var}[Y] \approx E\left[\left(\sum_{i=1}^N \left.\frac{\partial 
\mathcal{M}}{\partial x_i}\right|_{\mathbf{x}=\mathbf{x}_{\mu}} 
(x_i-x_{\mu,i})\right)^2\right]= \mathbf{J}\mathbf{\Sigma}\mathbf{J}^T,
\end{equation}
where $\mathbf{J}$ is the Jacobian row-vector $\mathbf{J} =
[\frac{\partial \mathcal{M}}{\partial x_1}, \frac{\partial
    \mathcal{M}}{\partial x_2}, \ldots,
  \frac{\partial\mathcal{M}}{\partial x_N}]$, and $\mathbf{\Sigma}$ is
the covariance matrix of the LECs. As mentioned, this is a relatively
computationally inexpensive method for estimating the variance of the
model prediction. It only requires the computation of $N=13$ partial
derivatives in the present case. The resulting central values as well
as the propagated LEC uncertainties are given in
\Tref{tab:model_uncertainties}.
\begin{table}[h!]
\center
\caption{Statistical uncertainties of the deuteron observables, the
  deuteron D-state probability $P_D$, and the $^1S_0$ effective range
  parameters that originate from the statistical uncertainty of the
  interaction. The energies, radii, effective range parameters, and
  electric moments are given in units of Mev, fm, fm, and $e$-fm,
  respectively. The $P_D$ is reported in $\%$. See the text for
  further details.}
\label{tab:model_uncertainties}
\lineup
\begin{tabular}{@{}llll}
\br
            Observable           & Theory & Experiment  & Ref. \\
\mr
\multicolumn{4}{c}{$^{1}S_0$ effective range}     \\
\mr
\multirow{2}{*} {$a^C_{pp}$} & \multirow{2}{*}{$\-7.811(1)$}  & $\-7.8196(26)$ &\cite{Ber88} \\
          &          & $\-7.8149(29)$ &\cite{San83} \\
\multirow{2}{*}{$r^C_{pp}$} & \multirow{2}{*}{2.754(2)}    &  2.790(14) &\cite{Ber88}  \\
          &          &  2.769(14) & \cite{San83} \\
a$_{np}$     & $\-23.74(17)$  & $\-23.740(20)$ & \cite{Machleidt2001}\\
r$_{np}$     & $ 2.683(2)$    & 2.77(5) & \cite{Machleidt2001}\\
a$_{nn}$    & $\-18.95$ &  $\-18.95(40)$ & \cite{Gon06,Chen08}\\
r$_{nn}$    & $ 2.79$  &  $2.75(11)$ & \cite{Miller90}\\
\mr
\multicolumn{4}{c}{$^2$H}\\
\mr
E$_{\rm gs}$ &  $\-2.222(8)$   &  $2.22456627(46)$ & \cite{CODATA2010} \\ 
$\langle r^2_m \rangle^{1/2}$ & $1.968(3)$ &  1.97535(85) & \cite{Huber98}\\ 
$P_D$ & $4.04(4)$  & --&\\ 
$Q_{\rm gs}$ & $0.272(1)$  & 0.2859(3)& \cite{David79,Ericson83}\\ 
\br
\end{tabular}
\end{table}
Overall, the experimental values are more precise than their
theoretical counterpart. Also, the predictions of the NNLO model are
not consistent with the experimental values in all cases. In
particular, the deuteron radius and quadrupole moment are both
underestimated. The discrepancies are not so severe, and we note that
the values that are presented here are almost identical to what is
obtained using high-precision interaction models\footnote{See for example the  CD-Bonn~\cite{Machleidt2001} interaction model.}. 
The ground state energy of the
deuteron has a relative theoretical error of 0.4\%. Thus, the
statistical uncertainty of the interaction generates an error that is
$10^4$ times greater than the experimental uncertainty.  Regarding the
uncertainties of the effective range parameters, they again reflect
the fact that the $pp$ scattering data is more precise than the $np$
data. Our model predictions for the $pp$ parameters are
consistent with the analysis of Ref.~\cite{San83}. We have also
determined the uncertainties for a selected set of partial-wave phase
shifts; see \Tref{tab:phase_shifts} and plotted in
\Fref{fig:phase_shifts}.
\begin{table}[h]
\center
\caption{Selected proton-proton and neutron-proton scattering phase shifts relevant at NNLO. Note that the potential has been optimized with respect to scattering data with $T_{\textnormal{LAB}}\leq125$ MeV. The remaining $np$ and $pp$ phase shifts have $<0.001$ degrees in statistical uncertainty. See the text for details.}
\footnotesize
\label{tab:phase_shifts}
\lineup
\begin{tabular}{@{}lrrrr|rrrrrrr}
\br
\tiny
$T_{\textnormal{LAB}}$ & pp$^1S_0$ &  pp$^3P_1$ &  pp$^3P_0$ &  pp$^3P_2$ & np$^1S_0$ & np$^3S_1$ & np$\varepsilon_1$ & np$^1P_1$ & np$^3P_1$ & np$^3P_0$ & np$^3P_2$ \\
\mr
1   & 32.80     & -0.08     & 0.14      & 0.014       & 62.05 & 147.74 & 0.11 & -0.19 & -0.11 & 0.18 & 0.02 \\                              
    & $\pm0.00$ & $\pm0.00$ & $\pm0.00$ & $\pm0.00$  & $\pm0.02$ & $\pm0.00$ & $\pm0.00$ & $\pm0.00$ & $\pm0.00$ & $\pm0.00$ & $\pm0.00$ \\
5   & 54.96 & -0.89 & 1.62 & 0.22                    & 63.58 & 118.15 & 0.68 & -1.54 & -0.92 & 1.66 & 0.25 \\                              
    & $\pm0.00$ & $\pm0.00$ & $\pm0.00$ & $\pm0.00$  & $\pm0.00$ & $\pm0.00$ & $\pm0.00$ & $\pm0.00$ & $\pm0.00$ & $\pm0.00$ & $\pm0.00$ \\
10  & 55.38 & -2.03 & 3.84 & 0.65                    & 59.87 & 102.56 & 1.17 & -3.16 & -2.02 & 3.74 & 0.71 \\                              
    & $\pm0.00$ & $\pm0.00$ & $\pm0.00$ & $\pm0.00$  & $\pm0.00$ & $\pm0.00$ & $\pm0.00$ & $\pm0.00$ & $\pm0.00$ & $\pm0.00$ & $\pm0.00$ \\
25  & 48.88 & -4.84 & 8.82 & 2.50                    & 50.67 & 80.45 & 1.82 & -6.59 & -4.77 & 8.36 & 2.58 \\                               
    & $\pm0.00$ & $\pm0.00$ & $\pm0.00$ & $\pm0.00$  & $\pm0.00$ & $\pm0.00$ & $\pm0.00$ & $\pm0.00$ & $\pm0.00$ & $\pm0.001$ & $\pm0.00$ \\
50  & 39.09 & -8.25 & 11.50 & 5.87                   & 39.95 & 62.28 & 2.18 & -10.00 & -8.18 & 10.72 & 5.94 \\                             
    & $\pm0.00$ & $\pm0.00$ & $\pm0.00$ & $\pm0.00$  & $\pm0.00$ & $\pm0.00$ & $\pm0.00$ & $\pm0.00$ & $\pm0.00$ & $\pm0.00$ & $\pm0.00$ \\
100 & 24.82 & -14.03 & 7.47 & 10.76                  & 25.28 & 41.90 & 2.56 & -14.39 & -14.06 & 6.50 & 10.75 \\                            
    & $\pm0.01$ & $\pm0.00$ & $\pm0.01$ & $\pm0.00$  & $\pm0.01$ & $\pm0.01$ & $\pm0.01$ & $\pm0.03$ & $\pm0.00$ & $\pm0.01$ & $\pm0.00$ \\
150 & 14.27 & -20.21 & -0.78 & 12.64                 & 14.71 & 28.40 & 2.81 & -18.09 & -20.36 & -1.80 & 12.53 \\                           
    & $\pm0.02$ & $\pm0.01$ & $\pm0.03$ & $\pm0.00$  & $\pm0.02$ & $\pm0.03$ & $\pm0.02$ & $\pm0.13$ & $\pm0.01$ & $\pm0.03$ & $\pm0.00$ \\
200 & 6.06 & -27.13 & -10.50 & 12.29                 & 6.59 & 17.97 & 2.82 & -21.60 & -27.38 & -11.53 & 12.12 \\                           
    & $\pm0.05$ & $\pm0.02$ & $\pm0.04$ & $\pm0.01$  & $\pm0.05$ & $\pm0.06$ & $\pm0.06$ & $\pm0.35$ & $\pm0.03$ & $\pm0.04$ & $\pm0.01$ \\
250 & -0.14 & -34.63 & -20.80 & 10.58                & 0.55 & 9.56 & 2.47 & -24.78 & -34.96 & -21.83 & 10.36 \\                            
    & $\pm0.10$ & $\pm0.05$ & $\pm0.06$ & $\pm0.01$  & $\pm0.11$ & $\pm0.10$ & $\pm0.13$ & $\pm0.74$ & $\pm0.05$ & $\pm0.06$ & $\pm0.01$ \\
300 & -4.25 & -42.35 & -31.21 & 8.16                 & -3.39 & 2.93 & 1.65 & -27.22 & -42.72 & -32.22 & 7.93 \\                            
    & $\pm0.18$ & $\pm0.08$ & $\pm0.10$ & $\pm0.02$  & $\pm0.19$ & $\pm0.16$ & $\pm0.22$ & $\pm1.36$ & $\pm0.09$ & $\pm0.10$ & $\pm0.02$ \\
350 & -6.17 & -49.76 & -41.23 & 5.56                 & -5.17 & -1.86 & 0.39 & -28.28 & -50.13 & -42.20 & 5.34 \\                           
    & $\pm0.28$ & $\pm0.14$ & $\pm0.14$ & $\pm0.02$  & $\pm0.28$ & $\pm0.23$ & $\pm0.32$ & $\pm2.20$ & $\pm0.15$ & $\pm0.15$ & $\pm0.02$ \\
\br
\normalsize
\end{tabular}
\end{table}
Not surprisingly, the uncertainties increase with increasing
$T_{\textnormal{LAB}}$. This is most likely due to the fact that the chiral
NNLO interaction is a low-momentum theory, and that the current
potential is optimized only to data with $T_{\textnormal{LAB}} \leq
125$ MeV. Thus, values beyond this cutoff are
extrapolations. Qualitatively, the results of a similar analysis using
the chiral two-pion exchange, coupled to a coarse-grained
representation of the short-ranged potential, exhibits a similar
pattern in the uncertainties.
\subsection{Sensitivity analysis of the model response}
In order to better understand the variance of the computed observables, we
quantify the relative importance of each input parameter in a
linearized sensitivity analysis. In general, an approximate 
expression for the response variance around a nominal value
$\mathbf{x}_{\mu}$, obtained by disregarding parameter correlations, is
given by
\begin{equation}
\textnormal{Var}[Y] \equiv \sigma_{Y}^2 \approx \sum_{i=1}^N \left(
\left.\frac{\partial \mathcal{M}}{\partial x_i}
\right|_{\mathbf{x}=\mathbf{x}_{\mu}}\right)^2 \sigma_{x_i}^2.
\end{equation}
This allows us to define the linearized relative importance, or
sensitivity, of parameter $x_i$ on the variance of the response $Y$ as
\begin{equation}
S_i = \left(\frac{\partial \mathcal{M}}{\partial x_i}\right)^2 \left(
\frac{\sigma_{x_i}}{\sigma_Y}\right)^2
\end{equation}
In the case of independent variables $x_i$, the sum of parameter
sensitivities is normalized to one 
(i.e., $\sum_{i=1}^{N}S_i=1$). However, for correlated input, as in our
case, the sum will not add up to unity. Still, we have carried out
this analysis for all deuteron observables that we have computed in
this paper, see \Tref{tab:sensitivities}. This analysis reveals the
relative importance of the various sources of uncertainty when the
structure of the nuclear wave function is taken into account. It is
clear that the largest component of the uncertainties for the deuteron
observables originate in the $\pi-N$ LECs $c_1,c_3,c_4$, at least for
the current optimum $\mathbf{x}_{\mu}$.
\begin{table}
\center
\caption{Sensitivity or importance $S_i$, in percent, of the different
  NNLO LECs with respect to the variance of the calculated deuteron
  properties. The values are normalized with respect to the
  uncorrelated uncertainties. A large number indicates that the LEC
  has a large impact on the uncertainty of the observable.}
\footnotesize
\label{tab:sensitivities}
\lineup
\begin{tabular}{@{}lllllll}
\br
Obs. & $c_1$&$c_3$&$c_4$&$\tilde{C}_{^3S_1}$&$C_{^3S_1}$&$C_{E_1}$ \\  
\mr
\multicolumn{7}{c}{$^2$H}     \\
\mr
   E$_{\rm gs}$                       &    28.0    &   8.1   &   46.9   &    5.8   &   4.8  &   6.4  \\
    $\langle r^2_{m} \rangle^{1/2}$   &    28.3    &   8.6   &   46.3   &    5.6   &   5.2  &   6.0  \\
   $P_D$                                &     8.5    &  3.3    &  70.4    &   1.7    &  1.9   & 14.3    \\
   $Q_{\rm gs}$                                 &    34.3    & 10.5    &  38.6    &   6.6    &  6.3   &  3.7   \\
\br
\end{tabular}
\end{table}

\section{Conclusions}
In conclusion, we have optimized a chiral NNLO two-body potential and
quantified the statistical uncertainties of the LECs. The relative
uncertainties are below 1\% except for $c_1$ and $C_{^1P_1}$ where it
is 5.84\% and 2.60\%, respectively. From a correlation analysis of the
parameters we observed that the two groups of $S-$wave and $P-$wave
contacts mostly anti-correlate. The groupwise correlation between
partial-waves of different angular momentum is consistent with similar
analyses of coarse-grained nuclear
potentials~\cite{Navarro14,Navarro13_1,Navarro13_2}.

The uncertainties of the LECs are reflected in a statistical
uncertainty of calculated deuteron observables, phase shifts, and
effective range parameters. In general, the relative uncertainty of
the theoretical results are roughly $0.5\%$, which is on the order of
the uncertainties of the LECs. Naturally, only the lightest nuclear
systems can be solved exactly using numerical methods. Before
proceeding with a similar analysis for heavier systems, it can be of
importance to extract the derivatives of the objective function with
higher numerical precision than what can be offered from
finite-difference approximations or a bivariate spline of the
objective function.  One way of obtaining the necessary derivatives to
machine precision is automatic differentiation~\cite{griewank2008edp},
an approach that we are actively pursuing.
Also, when using certain many-body methods or
employing various truncations for solving the Schr\"odinger equation,
the propagated statistical uncertainty could become conflated with the
uncertainties inherent to the numerical solution of the many-body
problem. Thus, for accurate uncertainty estimates of nuclear models it
is important to derive mathematically sound error
estimates of the nuclear many-body method itself. This is of course
relevant for an accurate extraction of the systematic uncertainty,
which is always a much more complicated source of
uncertainty~\cite{Witeks_guide}.

\ack This work was supported by the
Research Council of Norway under contract ISP-Fysikk/216699; by the
Office of Nuclear Physics, U.S.\ Department of Energy (Oak Ridge
National Laboratory), under grant nos.\ DE-FG02-03ER41270 (University
of Idaho), DE-FG02-96ER40963 (University of Tennessee), DE-AC02-06CH11357 
(Argonne), and DE-SC0008499 (NUCLEI SciDAC collaboration);
and by the European Research Council under the European Community's Seventh Framework
Programme (FP7/2007-2013)/ERC grant agreement no.\ 240603. This
research used computational resources of the Notur project in Norway and the National
Supercomputer Centre (NSC) at Linköping University provided by the Swedish
National Infrastructure for Computing (SNIC) .

\appendix
\setcounter{section}{1}
\section*{References}
\bibliography{jpg_uncertainty}
\bibliographystyle{unsrt}
\end{document}